# Integrating Python data analysis in an existing introductory laboratory course

Eugenio TUFINO (1), Stefano OSS (1), and Micol ALEMANI (2)

*(1) Department of Physics, University of Trento, 38123 Trento, Italy*

*(2) Institute of Physics and Astronomy, University of Potsdam, 14476 Potsdam, Germany*

**Abstract.** In this article we describe how we successfully incorporated data analysis in Python in a first-year laboratory course without significantly altering the course structure and without overburdening students. We show how we created and used carefully designed Jupyter Notebooks with exercises and physics application examples that allow students to master data analysis programming in the laboratory course. We use these Notebooks to guide students through the fundamentals of data handling and analysis in Python while performing simple experiments. We present our teaching approach and the developed materials. We discuss the effectiveness of our intervention based on the results from pre- and post-course questionnaires and students' group work. The results presented give insights about advantages and challenges of introducing computation at the early stage of the curriculum in a laboratory course setting and are informative for other instructors and the physics education research community.

**Introduction**

In the fast-paced era of technological advancement, computational skills have become a necessity across various disciplines, including physics. According to the World Economic Forum's projections for 2025 [1], there is a quickly rising demand in the job market for individuals who can harness the power of computation to analyse complex data and create innovative solutions to problems.

Also, in all fields of experimental science, the rate at which data is being generated is accelerating, and the use of robust tools for data analysis and interpretation has become a necessity. Computational skills are increasingly demanded especially in large-scale projects in particle physics or astrophysics, where both the management and analysis of big data are essential.

Due to these demands, there is a need to teach 'computational skills' in the physics curriculum, a step beyond the conventional use of spreadsheets and integrated mathematical computing packages typically found in physics laboratory courses. By fostering these competencies, students are not only prepared for their future careers in areas that increasingly require these skills but can also enrich their understanding of the studied subjects.

Introducing computation into high school or undergraduate courses has recently become an active area of physics education research. It is now widely recognized that in the practice of physics, there are three fundamental elements: experimentation, theory, and computation. This interest in integrating computation has led to specific recommendations from the American Association of Physics Teachers (AAPT) [2-4], and from the Partnership for Integration of Computation into Undergraduate Physics (PICUP) group [5].

Various initiatives have already been made to integrate computational elements into physics courses. For instance, work conducted at the University of Oslo, Norway, explored the use of "computational essays" into physics courses for improving communication and modeling across different subjects, such as mechanics, electricity, and magnetism [6]. For a comprehensive review of the integration of computational elements into physics education, see the following article and references within [7].

The above efforts are focused on computation as a tool to investigate physical concepts and applications in lectures. On the other hand, data analysis is a crucial skill for experimental physics. Physics laboratory courses (PLCs) offer a special and desirable environment for introducing programming skills early in the curriculum. In fact, in these settings students have many opportunities to practise programming data analysis with their own data and in an authentic setting. However, there are some challenges to be overcome in order to not overburden students in these settings. Laboratory courses are in fact already dense with learning goals, for example being able to apply statistical analysis techniques, deal with measurement uncertainties or document experiments [8].

As an effort to integrate computation at the early stages of the undergraduate curriculum, this article investigates how to teach data analysis programming in Python in an introductory physics laboratory course (PLC). The choice of Python is motivated by its extensive use in scientific research and its appeal in the job market.

While we are aware that there already exist PLCs that incorporate the use of Python, the contribution of this study lies in exploring a feasible way to reach this goal without significantly changing the course structure and without overburdening the students and to evaluate its effectiveness. To our knowledge, there are no studies that address this particular aspect and also that discuss the assessment of the effectiveness of the intervention.

In the sections that follow, we provide a detailed overview of the design, implementation, and evaluation of this Python data analysis introduction within a PLC.

**Design and implementation of the intervention**

To introduce Python programming language within the introductory PLC we utilised Jupyter Notebook (JN), an open-source web-browser-based interactive application that facilitates the creation and sharing of documents containing code, equations, figures, and text [9]. We developed several introductory JNs to teach the fundamentals of Python in presence, actively engaging students in PLCs in a collaborative learning process. Additionally, we prepared more JNs, with examples of physics applications and data handling, that students can use independently and asynchronously.

*Course structure*

We implemented the newly designed introduction to Python data analysis in the first-semester introductory laboratory course for physics major students at the University of Potsdam in Germany. This course is the first of a sequence of obligatory modules of the PLC taking place during the first four semesters of the curriculum. These modules have been recently restructured by one of the authors (M. A.) focusing on students' development of experimental skills such as design, modelling, communication, and technical skills [10]. The teaching approach abandons traditional cookbook-style experiments, in favour of a more student-centred authentic approach. One notable feature includes the use of laboratory notebooks rather than conventional laboratory reports, to promote a more engaged, authentic and reflective learning experience [11, 12]. Communication and collaboration are fostered through carefully designed activities conducted in groups. We notice here, that before the intervention on computation described here, students had already received instruction and practised on how to evaluate measurement uncertainties, how to conduct graphical analysis, and how to maintain a laboratory notebook.

*Integrating computation with Jupyter notebook in the lab course*

Our approach to computational instruction consisted in a gradual increase in difficulty, from the basics of programming (for example doing simple calculations and text formatting) to more advanced data-handling and analysis techniques. While developing the material, we made sure that no prior coding experience was required. Since we were able to base our data analysis programming activities on existing content knowledge on graphical analysis and measurement uncertainties, our JNs were designed to directly engage students in solving coding exercises and problems in groups. Examples of these exercises included calculating the mean and standard deviation, plotting data with error bars, performing linear regression, computing statistical quantities like $\chi^2$ and $r^2$, and creating histograms. We made extensive use of well-known scientific libraries such as Numpy, SciPy, and Pandas, integral to modern scientific computing.

To motivate students in engaging further with data analysis in Python, we provided them also with real-world scientific applications. For example, we showed how to import a data file and calculate the moving average for smoothing the data of the number of sunspots in time (see figure 1).

The JNs here described are accessible on the supplementary material and on GitHub (in the English and German versions) for broader educational use [13].

Instructors of similar courses are welcome to utilise these JNs, which cover standard topics for an introductory physics lab course. While the examples and practical scientific applications provided in our notebooks offer comprehensive perspectives, instructors can adapt the content to fit the specific requirements and objectives of their courses, for example, by introducing examples from biology, engineering, and other disciplines.

We notice here that JNs can be accessed through cloud platforms like Google Colab [14], allowing for cloud usage without software installation. However, due to privacy constraints in Germany, we were unable to fully exploit this cloud-based feature and instead worked within the Anaconda environment in the classroom [15]. Even if the Anaconda environment requires students to go through installation processes, we did not experience problems with this aspect.

In two dedicated laboratory sessions, we introduced the foundations of Python for basic data analysis using four JNs (see table 1). Additional JNs were provided for students to work asynchronously on application examples to further deepen their skills. In table 1, we describe the details of the intervention.

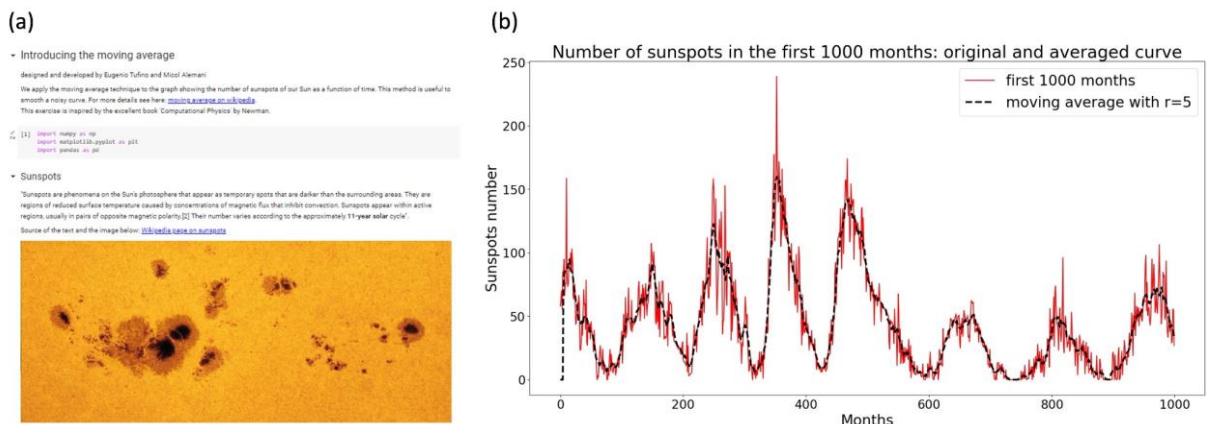

**Figure 1.** (a) Snapshot of a Jupyter Notebook containing text, code and images. The goal of this notebook is to introduce the use of the moving average for smoothing noisy curves. An example of the use of the moving average is shown in (b) for the number of sunspots as a function of time.

| Notebooks's Material | Short description | Type of instruction |
|---|---|---|
| Getting Started with Python (four JNs) | First laboratory session: Introduction to various JN platforms and their installation. Basic calculations, formatting text, Latex, basic plot.<br>Second laboratory session: Introduction of scientific libraries, use of Pandas dataframe, linear fit, statistical quantities like $\chi^2$ and $r^2$, and histograms | Two in-person lab sessions, three hours each |
| Application examples (five JNs) | Implementing Python to analyse real-world physics phenomena: Calculating the moving average applied to sunspot time-series, plotting and analysing light source illuminance vs. distance, creating histograms of the period of oscillation of a pendulum, performing a simulation of the parabolic motion, and plotting atmospheric $CO_2$ data from website | Asynchronous work, to be done independently for the next lab module |
| Additional techniques (two JNs) | Techniques for importing and manipulating data files across various platforms like Jupyter Lab-Anaconda, Google Colab, and the internet | Asynchronous work, to be done independently for the next lab module |

**Table 1.** Python data analysis curriculum implemented in the introductory laboratory course.

**Methods and Data Collection**

To evaluate the impact of our intervention, we administered pre- and post-course questionnaires in German. The pre-survey served as a valuable tool in assessing students' previous experience with programming and their views on computational skills in general, allowing the instructors to tailor the course content and approach accordingly. The post-survey provided us with students' feedback on the intervention.
In addition to the questionnaires, we have also evaluated hands-on exercises and have performed an in-depth analysis of students' work.
This multi-faced methodology allowed us to understand students' prior competencies, their engagement with the computational components, and their reception of the new methods, as well as to assess the development of their skills throughout the course.

Participants: 48 students participated voluntarily in this study. During the course students were supervised by teaching assistants (TAs), and by authors of the study, one of which is also the course coordinator. Students worked typically in groups of three. To form a group, students had the option to choose their own group members. If they did not express a preference, they were assigned by the instructor. For the assessment, students submitted their group-work.

Pre- and Post-surveys: The questionnaires were designed and validated by our team and filled out by students in a paper format. All participants were informed of the study's purpose and participated voluntarily. To be able to match students' answers to the questionnaires, students were asked to self-generate an anonymous code. The pre- and -post surveys were compiled by students individually. We obtained 42 matched answers between the pre-course and post-course surveys.

The pre-course survey consisted of 11 questions (see supplementary material for the surveys) and was designed to investigate students' initial background and familiarity with programming. It also sought to understand students' expectations and attitudes towards the integration of programming into the first semester PLC. The post-course survey consisted of an expanded set of questions. To be able to do a comparison, some of the pre-survey questions on students' expectations regarding programming were repurposed in the post-survey. In the post-survey students were also given the opportunity to articulate their thoughts through open-ended answers, express their judgments on specific aspects of the Python module, and provide feedback on group work. The inclusion of open-ended questions allowed for a more in-depth understanding of students' engagement in the course.

In the analysis of the open-ended responses, we employed an inductive methodology, utilising coding techniques to identify patterns and themes [16].

Analysis of Students' Work during the first semester: We examined how students tackled in groups the exercises given during the two lessons in the first semester fully dedicated to Python. Furthermore, we analysed laboratory notebooks of student groups during the two subsequent experiments of the first semester, specifically looking at how they utilised Python and the JNs. This analysis provided insight into the students' practical application of programming skills and the integration of computational methods into their experimental work.

Analysis of Students' Work during the following (second) semester: At the beginning of the second semester, we let student groups solve two in class exercises. These were specifically designed to challenge students to apply the techniques they had learned during the previous semesters and to reinforce their understanding of the programming concepts. A description of these exercises is in table 2.

| Exercise | Objectives | Methods to use |
| --- | --- | --- |
| Exercise 1: Methane Annual Increase | Analyse global atmospheric methane increase using data from the Global Monitoring Laboratory website | Import data from webpage. Plot the data and uncertainties |
| Exercise 2: Is the Stefan-Boltzmann Law Applicable to a Light Bulb? | Students are tasked to investigate whether the Stefan-Boltzmann law holds true for a light bulb by analysing measured data | Import voltage, current, and temperature (T) data from a provided Excel file. Plot electric power versus $T^4$. Then, perform linear regression and calculate $r^2$ |

**Table 2.** Description of the two exercises assigned to student groups at the start of second semester.

Further insights were gained from observing how well students used Python for data analysis and graphing techniques in their second semester experiments.

**Results and Discussions**

*Pre-survey Findings*

From the pre-survey, we gathered information about students' previous experience in programming and on their views and expectations on learning how to do data analysis in Python in the PLC. Regarding previous programming abilities, 62% of the students already had some experience (ranging from a few weeks to several years), with 38% of those students being somehow familiar with Python and only a few (7%) having extensive experience with JNs.
We also obtained important insights regarding students' views on learning programming: Students unanimously recognize the importance of learning programming at an early stage of their undergraduate education and for their careers.
For detailed response percentages to questions about the significance of programming in the 21st century, its relevance to physicists, and the value of introducing programming in the first-year physics lab, please refer to the stacked bar chart in figure 2.

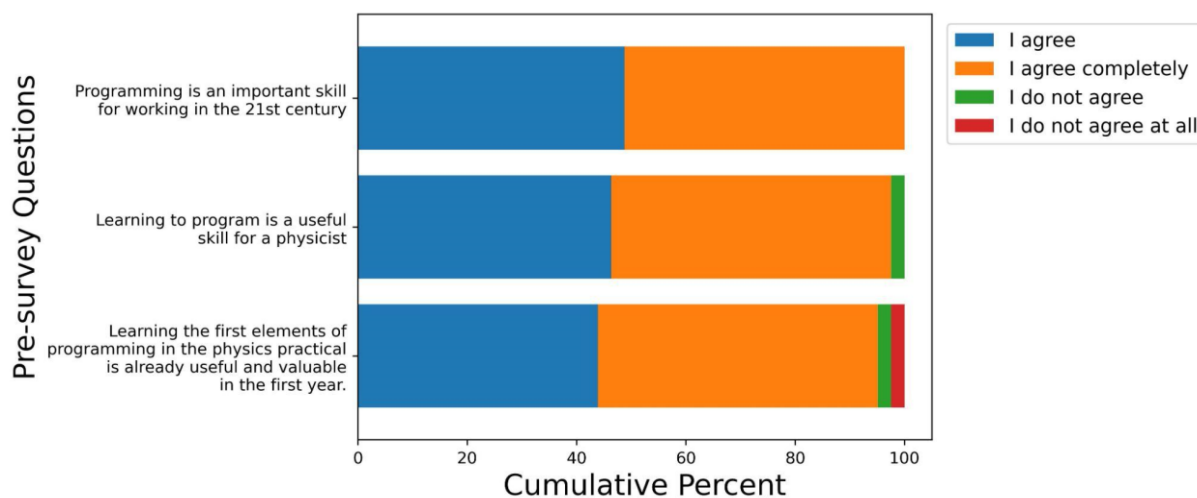

**Figure 2.** Stacked bar plots of students' answers to pre-survey questions about the importance of programming.

In our opinion, these findings show that students have very positive attitudes towards programming. This result can potentially encourage other instructors to include computational aspects in their PLC, since students' epistemology on a certain subject clearly influences their learning [17]. Moreover, instructors do not have to overcome possible students' resistances to the introduction of new subjects.

*Analysis of Students' Work during the first Semester*

In the introductory sessions dedicated to Python, students were guided through various exercises contained in the JNs we provided (see table 1 and supplementary material for details). Overall, the sessions went smoothly. While some groups finished well ahead of schedule, others completed the exercises within the allotted time. At the end of each session, student groups were required to submit their completed JNs. They have been graded by the TAs by using a rubric. No significant difficulties were observed.
In the two laboratory sessions following the two introductory sessions dedicated to Python, students worked in groups on two experiments. In the first experiment, inspired by N. Holmes [18], students brought elastic objects from home to explore whether they obey Hooke's law. In the second experiment, students worked with a pendulum.

Examples of the student group's works using Python during the first experiment are shown in figure 3. In the first example (see figure 3(a)) students studied the behaviour of an elastic band brought from home, attempting using Python to determine which fit was most suitable to describe the data. At the end of this laboratory session, students had also the opportunity to freely develop their own "research question" on the studied system. For instance, some groups studied what happens as the elastic band approaches its breaking point. An example of such inquiry can be seen in figure 3(b). After breaking a couple of thin rubber bands, this group of students used a thicker rubber band and attached several weights on it. By doing a graphical analysis in Python of the rubber band elongation as a function of the attached mass, they realised the existence of two regimes in the graph. They therefore did two different fits of the data. Notice that in their graph, they also showed (in orange) the data obtained by removing the weights from the rubber band, indicating the existence of a hysteresis in the system.

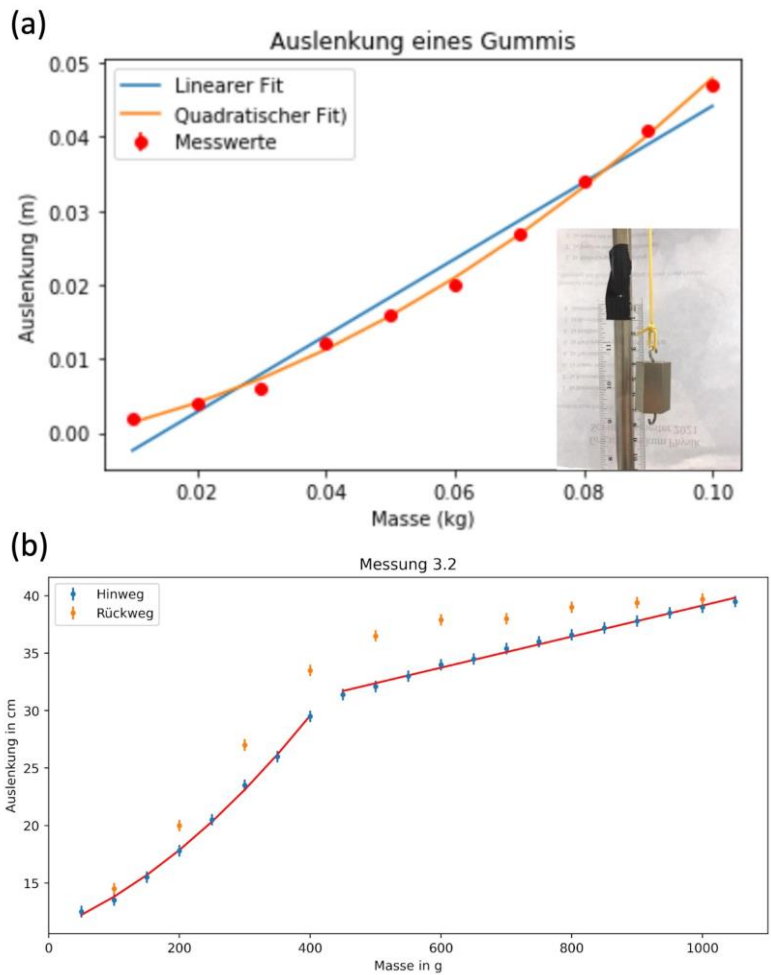

**Figure 3.** Extracts from laboratory notebooks of two student groups. (a) Graphical analysis in Python of the dependence between a rubber band elongation (in German 'Auslenkung') and the attached mass (in German 'Masse') with linear and quadratic fit. In the inset is a picture of the experimental setup. (b) Another plot of data obtained with a set-up as in (a) in which students analyse the behaviour of a rubber band near its breaking point (see text).

As exemplary shown in figure 3, during our examination of students' laboratory notebooks of these two experiments, we observed a marked increase in the utilisation of Python and JNs. Even if the use of Python and JN was not mandatory, 83% of student groups used Python and JNs for their data analysis in the first experiment and 100% in the second experiment. These results indicate an increase in students' confidence and proficiency in using these tools. We also found that while students consider JNs effective for analysis and visualisation, they found them less convenient for taking real-time notes during the experimental process. Many (28% of the groups for both experiments) used the JNs only for the data analysis and preferred to handwrite the experimental details and documentation on tablets, paper, or use word processing software like Word, rather than recording them directly within the JNs.

Interestingly we observed an unexpected positive effect on students' attitudes during experimentation that might be related to the introduction of computation. We experienced a self-driven engagement of students in doing data analysis while (or immediately after) taking experimental data and not postponing the data analysis to the end of (or after) the lab. In past courses, we repeatedly encountered the attitude of postponing data analysis and had to openly discourage it in our PLC. Expert experimental physicists in fact use preliminary data analysis 'on the spot' to be able to progressively make decisions in the laboratory based on their experimental results and promptly recognise problems. Even if further effort must be spent to investigate this qualitative observation, we can speculate here that introducing computational aspects for the data analysis might make students want to engage in it more actively and promptly during the laboratory. They might consider data analysis as an integral part of an experimental challenge and therefore become more motivated in doing it immediately. As a positive possible consequence, they might also experience the importance of data analysis for decision making when they for example find unexpected results from their analysis and obtain more 'expert-like' attitudes toward experimental physics.

*Post survey results*

It is remarkable that students' already positive views on the importance of computational skills not only remained consistent but even improved between the pre-survey and the post-survey (see figure 2). It is well known by physics education research, such consistency is not a given, as students often approach new subjects with initial beliefs that may or may not align with their views after instruction.

By analysing the open-ended answers to the question "Why is it important to learn the first elements of programming in the physics lab course already in the first year?", we found that students perceived the utility of programming for data analysis in different settings: in the laboratory, in subsequent physics courses, and in their future workplace. They recognise the importance of an early exposure to these skills because it allows for more practice. When asked about the advantages of programming data analysis in Python, students mentioned the potential time-saving benefits through automation and the general applicability in subsequent physics courses and professional work. We believe that these students' responses collectively reflect a future-oriented mindset.

Students' evaluation on the course (see figure 4), revealed that the majority of students (65%) found the introduction to Python feasible, 17.5% believed it was 'Easy' or 'Very easy' and 7.5% of students that it was 'Difficult'. Notably, no students chose 'Very difficult'.

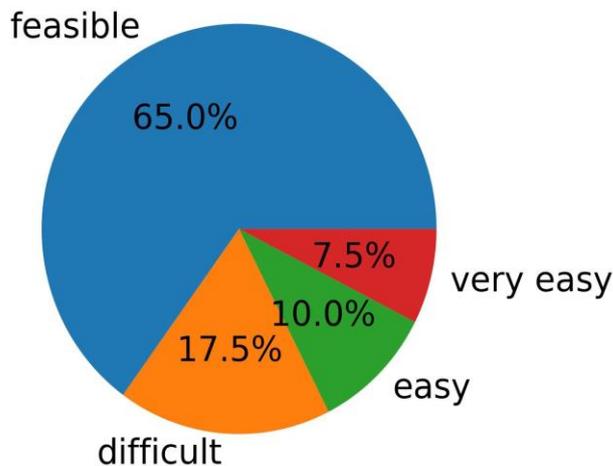

**Figure 4.** Pie chart of students' responses to a post-survey question about the feasibility of the Python introduction.

Those who found the introduction to Python 'Easy' or 'Very easy' generally had prior programming experience and appreciated the clear instructional structure. In the 'Feasible' category, 14 out of 25 students reported their feeling of a growth in their learning.

Examples of students' quotes (translated from German) explaining this growth are: '*If you have never worked with Jupyter Notebook before it is a bit complicated i.e. lots of new information. You have to develop an understanding of it first then it is easy*' and '*There were slight difficulties but if I tried hard it was doable*'.

Among the students that reported that the introduction to Jupyter was 'Difficult', some cited the high workload in the first year due to mathematics courses and physics, some reported their lack of prior experience as a challenge, and few complained about too much content and too little time. Since the large majority of the students (82.5%) did not experience difficulties during the course and nobody experienced the Python introduction as very difficult, the fact that 17.5% of the students encountered some challenges can be considered as normal in an educational setting.

The answers to further post-survey questions (see figure 5) allow for deeper investigation, revealing the following trends: A significant majority of students (84%) found the activities with JNs interesting, with 78% of students indicating plans to use JNs in the future and 86% plans to further deepen their skills in Python. However, we found that only 50% of the respondents felt capable of successfully using JNs for data analysis, while 29% were undecided or did not feel capable (21%).

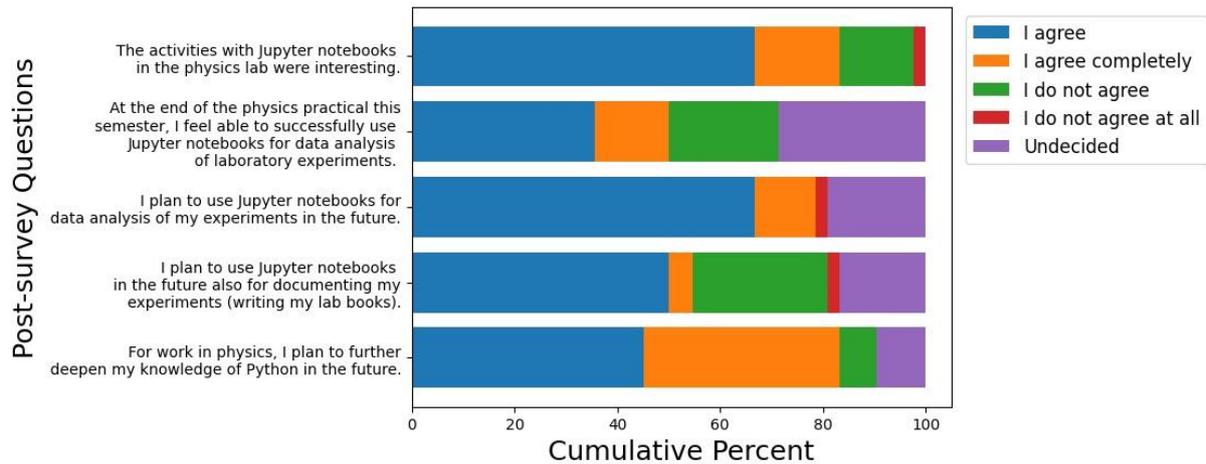

**Figure 5.** Stacked bar plots representing students' responses to five key questions of the post survey.

To better understand, we cross-referenced the responses from the seven students who reported difficulties (figure 4) with their answers to the questions in figure 5. We notice that despite the difficulties these 7 students did not feel overwhelmed and remained motivated, in fact most of them plan to further use JNs and want to deepen their skills in Python.

Interestingly, the survey responses from individual students partially contrast with our analysis of students' group work that shows all groups capable of successfully programming data analysis in Python, as discussed in the previous section. This discrepancy could potentially indicate that students are underestimating their own capabilities but could also be a result of group dynamics. To further examine this aspect, two additional questions in the post-survey were especially insightful. When asked, 'How satisfied are you with the group work during the Jupyter Notebooks activities?' the majority (71%) of students expressed satisfaction, while 12% were unsatisfied and another 12% neutral.

Examples of positive comments in the other (open ended) question are in the following (translated from German):

*"Was fun and effective too."," It was fun and educational","Worked well, everyone contributed to the task, was much more fun in the group", "We harmonised well as a group and everyone had their tasks","6 eyes see more than 2","You could always ask group members for help. I could not have done it alone","Because we worked very well together in the group"*

These results show that the teamwork was a strength in the trial, leading to good results.

On the other hand, students unsatisfied by group work mentioned:

*"All group members were at different levels, difficult distribution of tasks", "I had programming experience, had to wait", "The division of labour was not balanced, there were often disagreements", "Due to a group member, one had to take over the work for two people," , "The work often stuck to one person"*

This shows that targeted strategies on group work are needed in later iterations of the course, to balance groups equally, to make even more students comfortable.

The lack of confidence in using Python expressed by 21% of the students during the post-survey is not surprising. In fact, the post-survey was proposed after only four weeks of practice since the intervention began. As it is well known, the development of scientific skills takes several weeks. For example, in the case of scientific abilities in the ISLE environment, a period of five to eight weeks was measured [19].

To account for this, additional practice opportunities should be provided. In the following section we explain how we implemented this in our intervention.

*Students' Work in the following Semester*

At the beginning of the second semester, we gave students two exercises as review to Python programming (see methods section). The analysis of students' group work indicates a satisfactory students' mastery of the skills and techniques introduced through the JNs in the previous semester. In fact 83% of the group's work with Python data analysis was evaluated as good (50%) or excellent (33%). Only 17% of the groups needed improvements in their work, but no group work was evaluated as unsatisfying.

Additionally, the analysis of laboratory notebooks from student groups working on various experiments during the second semester also supports these findings. For example, some groups worked on an experiment focused on Brownian motion (see figure 6 for an extract from a student group work[1]) and demonstrated solid computational skills in analysing the data. Similar capabilities were exhibited by other groups in a different experiment.

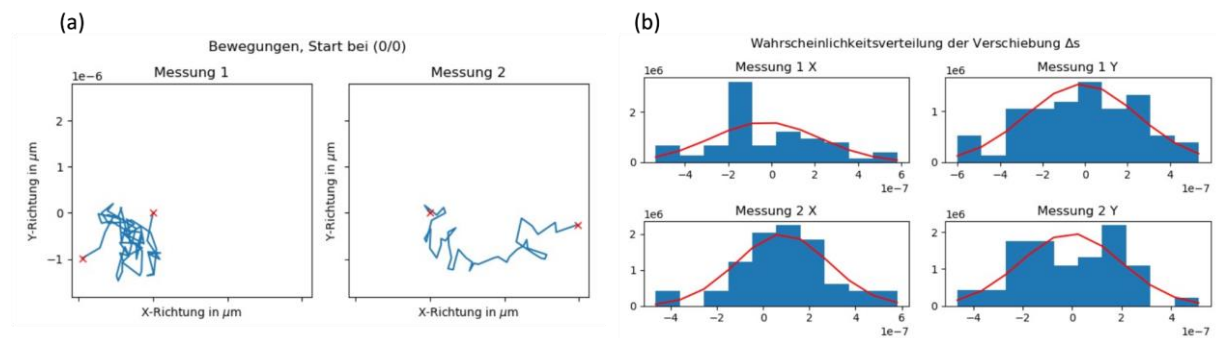

**Figure 6.** Extract from a student group's work on Brownian motion of a polystyrene particle in water. In (a) the experimental particle trajectories from two data sets obtained through video analysis of microscope images. In (b) students plot the relative frequencies of the particle's displacements divided by the bin size as a function of the particle's displacements.

**Conclusions**

In this study, we successfully integrated computational aspects into an existing physics laboratory course, assuring that students could learn a programming language without feeling overwhelmed, while keeping the course structure largely unchanged. The introduction of carefully designed JNs, not only fostered the acquisition of programming skills that students can use in their future courses and careers, but also helped the development of expert-like skills in data analysis. Instead of depending on ready-made data analysis tools as we did in the past, which often hinder the finer details, students are now adapting and implementing data analysis techniques in Python from scratch, tailored to the specific requirements of their experiments.

---

[1] As can be seen from the original excerpt, some of the graphs lack either axis description or scale. This behaviour, as is well known, is commonly found in PLC, and the use of Jupyter Notebooks has not impacted this habit.

The intervention was well-received, with students maintaining positive attitudes into the following semester. Even with diverse backgrounds in programming, almost all students recognized in the pre- and post- questionnaires the importance of learning programming for data analysis and positively valued the early integration into the curriculum. We consider this aspect as very positive since the students' epistemology has an important role in learning [17]. In fact, when introducing research-based changes or innovations in a course students' resistance to change can be a challenge [22]. We believe that the positive attitude we found among our students helped to incentivize them to acquire computational skills. Moreover, we found that students continued to meet other existing course goals effectively [9], as measured with the E-CLASS instrument [21-22].

However, it is important to acknowledge some limitations in our study. The primary limitation being that our analysis of students' computational skills focused on group work, and therefore might not provide a comprehensive understanding of individual student outcomes, but only on the class as a whole. Additionally, given that the number of students considered in this study is small (N<50), the statistical analysis is only descriptive and not inferential. As such, the conclusions may not apply in other contexts. Inf act, our study focused on a specific class of first-year physics students, some of whom had a background in programming, which has become more common in recent years in German schools. In another country, the proportion of students with such a background might vary, although initiatives introducing programming in high school have become widespread globally.

For the future, we plan to monitor students further by collecting more data and analyse the long-term effects and benefits of our intervention. We are also aware that there is a need to reinforce the computational skills acquired in our introduction by providing further opportunities for practice in subsequent laboratory courses. In circumstances where several faculty instructors are involved, collaboration is required.

We also aim to introduce formative assessment moments tailored for individual students. The next implementation, under consideration, will be proposed focusing possibly on a single environment to facilitate the process. In conclusion, our findings indicate that it is both feasible and beneficial to integrate Python-based data analysis in a PLC improving students' computational skills without significantly altering the course structure.

**Acknowledgments**

We gratefully thank A. Faber and R. Reimann for helping with the English-German translations of the Notebooks. M.A. thanks R. Bausinger for introducing her to the use of Jupyter Notebooks and E.T. thanks Giovanni Organtini for introducing him to the use of Pandas.